\documentclass[a4paper,11pt]{article}
\usepackage{pos}
\usepackage{subfigure}
\DeclareMathOperator{\Tr}{Tr}

\title{Lattice study of the confinement/deconfinement transition in rotating gluodynamics}
\ShortTitle{Lattice study of the confinement/deconfinement transition in rotating ...}

\author[a]{V.V.~Braguta}
\author[a]{A.Yu.~Kotov}
\author[b]{D.D.~Kuznedelev}
\author*[a]{A.A.~Roenko}

\affiliation[a]{ Bogoliubov Laboratory of Theoretical Physics, Joint Institute for Nuclear Research, Dubna, 141980 Russia }
\affiliation[b]{ Moscow Institute of Physics and Technology, Dolgoprudny, 141700 Russia }

\emailAdd{vvbraguta@theor.jinr.ru}
\emailAdd{kotov.andrey.yu@gmail.com}
\emailAdd{scope.denis@mail.ru}
\emailAdd{roenko@theor.jinr.ru}

\abstract{
We study the influence of relativistic rotation on the confinement/deconfinement phase transition in gluodynamics by means of lattice simulations. The simulation is performed in the reference frame which rotates with the system under investigation, where rotation is reduced to external gravitational field. The Polyakov loop and its susceptibility are calculated for various lattice parameters and values of angular velocities which are characteristic for heavy-ion collision experiments. Different types of boundary conditions (open, periodic, Dirichlet) are imposed in directions, orthogonal to rotation axis.
It is shown, that the critical temperature of the confinement/deconfinement transition in gluodynamics grows quadratically with  increasing angular velocity.
This conclusion does not depend on the boundary conditions used in our study and we believe that this is universal property of gluodynamics.
We also present first results of the study of the phase diagram of rotating QCD matter with fermions.
The results indicate, that effect of the rotation on fermions is opposite to gluons: it leads to the decrease of the critical temperature.
}

\FullConference{%
 The 38th International Symposium on Lattice Field Theory, LATTICE2021
  26th-30th July, 2021
  Zoom/Gather@Massachusetts Institute of Technology
}

\begin{document}
\maketitle

\section{Introduction}
A droplet of strongly interacting matter with nonzero angular momentum is expected to be created in non-central heavy ion collisions~\cite{Jiang:2016woz, Becattini:2007sr, Baznat:2013zx, STAR:2017ckg}. The resulting angular velocity of quark-gluon plasma can be determined from the polarization of $\Lambda$ hyperons and is expected to be $\Omega\sim6$ MeV~\cite{STAR:2017ckg}. Hydrodynamical simulations predict even larger values $\Omega\sim (20-40)$ MeV~\cite{Jiang:2016woz}. Such large values of angular velocity imply relativistic rotation and might lead to non-trivial phenomena based on the interplay between strong interaction and rotational motion. Well-known examples of such phenomena include chiral vortical effect~\cite{Vilenkin:1979ui, Kharzeev:2015znc, Prokhorov:2018qhq, Prokhorov:2018bql} and polarization of various particles~\cite{Rogachevsky:2010ys, Teryaev:2017wlm}.

There are a plenty of studies of QCD with nonzero rotation within effective models~\cite{Ebihara:2016fwa, Chernodub:2016kxh, Jiang:2016wvv, Zhang:2018ome, Wang:2018sur, Chernodub:2020qah, Chen:2020ath, Fujimoto:2021xix, Golubtsova:2021agl, Sadooghi:2021upd}.
Most of these studies are based on Nambu--Jona--Lasinio model, where the influence of nonzero rotation on gluonic degrees of freedom is often ignored.
Within this model the critical temperature in rotating QCD decreases due to suppression of the chiral condensate~\cite{Jiang:2016wvv}.
Other phenomenological approaches predict the same behaviour of the confinement-deconfinement critical temperature with angular velocity~\cite{Fujimoto:2021xix, Chen:2020ath, Chernodub:2020qah}. 
But in QCD the rotation acts on both fermions and gluons, and combination of these effects may lead to unexpected results.

In this Proceeding we present the lattice investigation of the phase diagram of gluodynamics with nonzero rotation. In particular, we concentrate on the confinement-deconfinement phase transition at finite temperature and how it is affected by relativistic rotation. In the end we also present the first preliminary results for the phase diagram of the  rotating QCD with fermions.

First results of this investigation were presented in~\cite{Braguta:2020biu} and for more detailed discussion we refer to~\cite{Braguta:2021jgn}.

\section{Rotating reference frame and temperature}
In order to carry out the study of the rotating system, we are going to change the reference frame to the one, which rotates with the system. In this reference frame rotation is reduced to an external gravitational field $g_{\mu\nu}$: 
\begin{equation}\label{eq:metric}
g_{\mu \nu} = 
\begin{pmatrix}
1 - r^2 \Omega^2 & \Omega y & -\Omega x & 0 \\
\Omega y & -1 & 0 & 0  \\ 
-\Omega x & 0 & -1 & 0 \\
0 & 0 & 0 & -1
\end{pmatrix}\, ,
\end{equation}
where $r=\sqrt{x^2+y^2}$ is the distance to the rotational axis $z$. Using this gravitational field, one can express the dynamics of the system by the Hamiltonian $H=\int dV \sqrt{g_{00}}\epsilon(\vec{r})$, where $\epsilon(\vec{r})$ is the energy density of the system. The partition function of the system under investigation is $Z=\Tr\exp[-\beta \hat{H}]$. Using standard techniques it can be rewritten as a path integral in Euclidean space $Z=\int DA \exp(-S_G)$ with the following expression for the action of gluodynamics:
\begin{equation}
S_{G} = \frac{1}{4 g^{2}} \int\! d^{4} x\, \sqrt{g_E}\,  g_E^{\mu \nu} g_E^{\alpha \beta} F_{\mu \alpha}^{a} F_{\nu \beta}^{a} \, .
\label{eq:gluometrics}
\end{equation}

Now it is worth mentioning {\it Ehrenfest-Tolman} effect: in external gravitational field the temperature is not constant in space in thermal equilibrium, the product $T(r)\sqrt{g_{00}}$ does not depend on spatial coordinates. In the case of external metrics, given by Eq.~\eqref{eq:metric}, $T(r)\sqrt{1-r^2\Omega^2}$ is constant. In other words, rotation effectively warms up the periphery of the volume: $T(r>0)>T(r=0)$.

Substituting the expression for the metrics in the Euclidean action~\eqref{eq:gluometrics}, one gets:
\begin{multline}\label{eq:rot_action}
	S_{G} = \frac{1}{2 g^{2}} \int\! d^{4} x \
    \big[(1 - r^2 \Omega^2) F^a_{x y} F^a_{x y} 
    + (1 - y^2 \Omega^2) F^a_{x z} F^a_{x z} + (1 - x^2 \Omega^2) F^a_{y z} F^a_{y z}
    + F^a_{x \tau} F^a_{x \tau} + F^a_{y \tau} F^a_{y \tau} +{} \\ {}+
    F^a_{z \tau} F^a_{z \tau} 	- 2 i y \Omega (F^a_{x y} F^a_{y \tau} + F^a_{x z} F^a_{z \tau})  
    + 2 i x \Omega (F^a_{y x} F^a_{x \tau} + F^a_{y z} F^a_{z \tau}) -
    2 x y \Omega^2 F^a_{x z} F^a_{z y}\big]\, .
\end{multline}

Note that action~\eqref{eq:rot_action} has the sign problem. In order to overcome the sign problem we perform simulations at imaginary angular velocity $\Omega_I=i\Omega$ and then perform an analytical continuation of our results $\Omega_I\to \Omega$. The final discretized version of the action is rather lengthy and can be found in \cite{Yamamoto:2013zwa,Braguta:2021jgn}. 

Simulations are performed with the lattice sizes $N_t\times N_z\times N_s^2$.
Lattice size $N_s$ in $x$- and $y$- directions is limited by the restriction $\Omega r<1$, thus one cannot perform simulations at arbitrary large lattices. It means, that one should treat carefully boundary conditions in these directions. In our study we introduced three types of boundary conditions:
\begin{itemize}
    \item {\it Open boundary conditions}. For these boundary conditions the sum in the action~\eqref{eq:rot_action} is restricted by plaquettes and 6-link elements, which lie inside the studied volume. All links outside the studied volume are ignored and are not taken into account in the action.
    \item {\it Periodic boundary conditions}. It is the standard type of boundary conditions used in the lattice simulation. It is incompatible with the velocity distribution of the rotating cube, however we used this type of boundary conditions in order to check robustness of our results.
    \item {\it Dirichlet boundary conditions}. In this type of boundary conditions we keep gauge field at the boundary fixed by setting all boundary links to unit matrix: $U_{x,\mu} = \hat 1$.
    Note that this type of boundary conditions explicitly violates $\mathbb{Z}_3$ center symmetry. 
\end{itemize}
It was shown, that in large volume limit the impact of non-periodic boundary conditions wanes, and only several layers near the boundary are affected by them~\cite{Braguta:2021jgn}.
In $z$- and $t$- directions we always keep periodic boundary conditions. 

In simulations the Polyakov loop, which is an order parameter for the confinement-deconfinement phase transition, and its susceptibility were calculated.

\section{Results for gluodynamics}
\begin{figure*}[htb]
\subfigure[]{\label{fig:O-Om-pl}
\includegraphics[width=.48\textwidth]{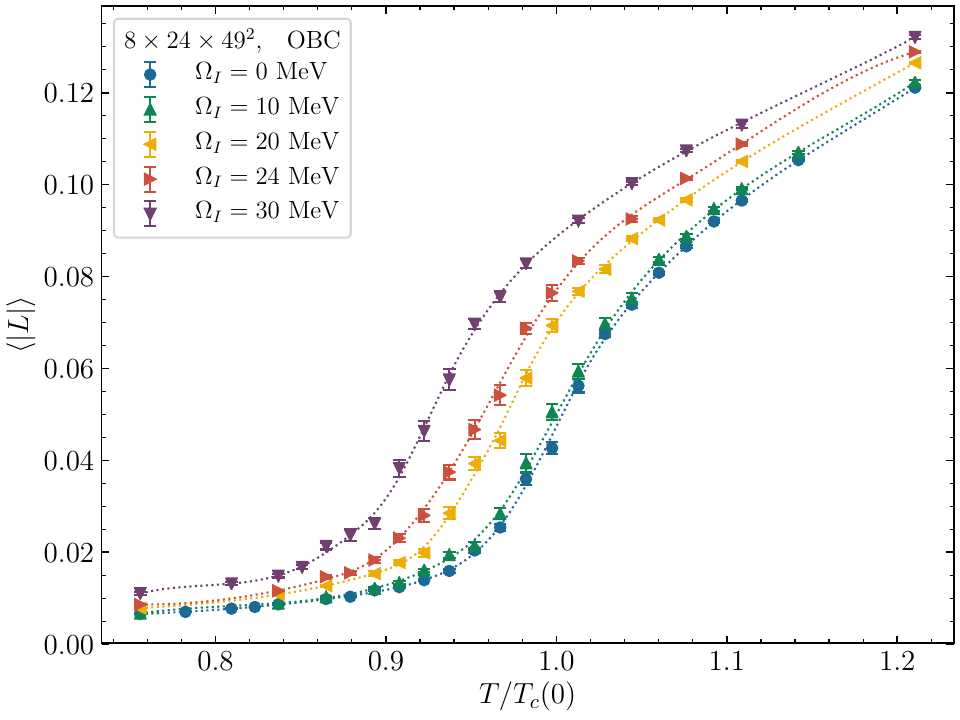}
}
\hfill
\subfigure[]{\label{fig:O-Om-chi}
\includegraphics[width=.48\textwidth]{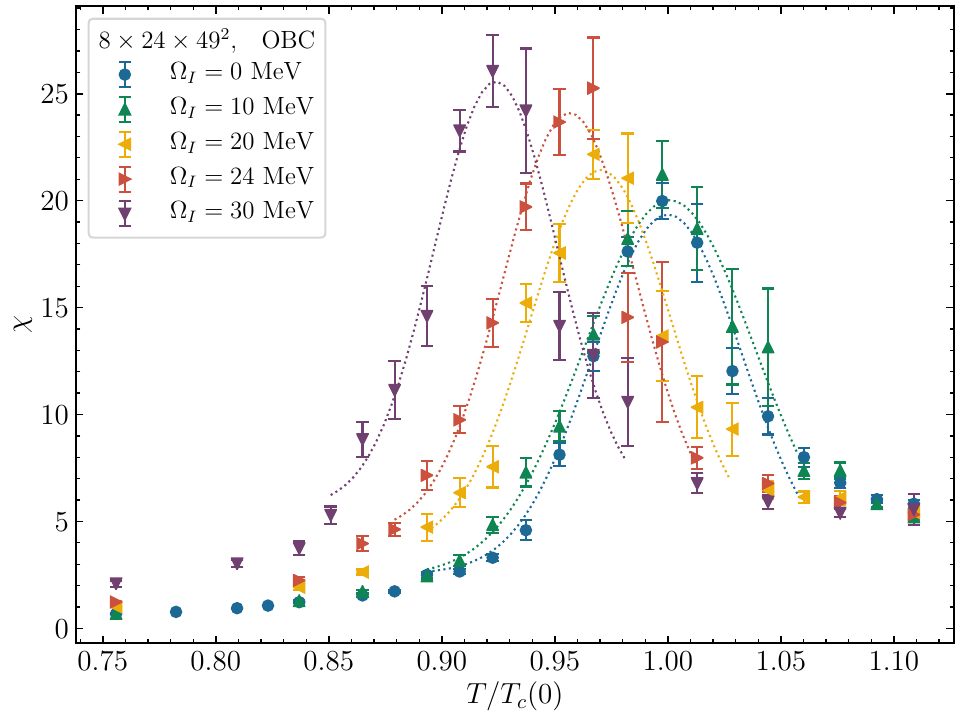}
}
\caption{The Polyakov loop~\subref{fig:O-Om-pl} and the Polyakov loop susceptibility~\subref{fig:O-Om-chi} as a function of temperature for different values of imaginary angular velocity $\Omega_I$. The results are obtained on the lattice $8\times 24\times 49^2$ with OBC. The lines for the Polyakov loop~\subref{fig:O-Om-pl} are drawn to guide the eye. The Polyakov loop susceptibilities~\subref{fig:O-Om-chi} are fitted in the vicinity of the phase transition by a Gaussian function.
Fig. from~\cite{Braguta:2021jgn}.}\label{fig:O-Om}
\end{figure*}

We believe that open boundary conditions (OBC) are the most appropriate for the description of the rotating medium, for this reason we present the detailed results of the simulation with this type of boundary conditions. In Fig.~\ref{fig:O-Om} the Polyakov loop and the Polyakov loop susceptibility are shown as functions of temperature $T/T_c(\Omega_I=0)$ for various values of (imaginary) angular velocity $\Omega_I$ for one particular lattice size $8\times 24\times 49^2$. Critical temperature can be determined from the inflection point of the Polyakov loop or from the peak of the susceptibility. It is clearly seen, that when one increases imaginary angular velocity $\Omega_I$, the critical temperature decreases. In order to make quantitative predictions, we fit the Polyakov loop susceptibility with a Gaussian function in the vicinity of the phase transition.

\begin{figure}[htb]
\subfigure[]{\label{fig:Tc-OBC-Om}
\includegraphics[width=.48\textwidth]{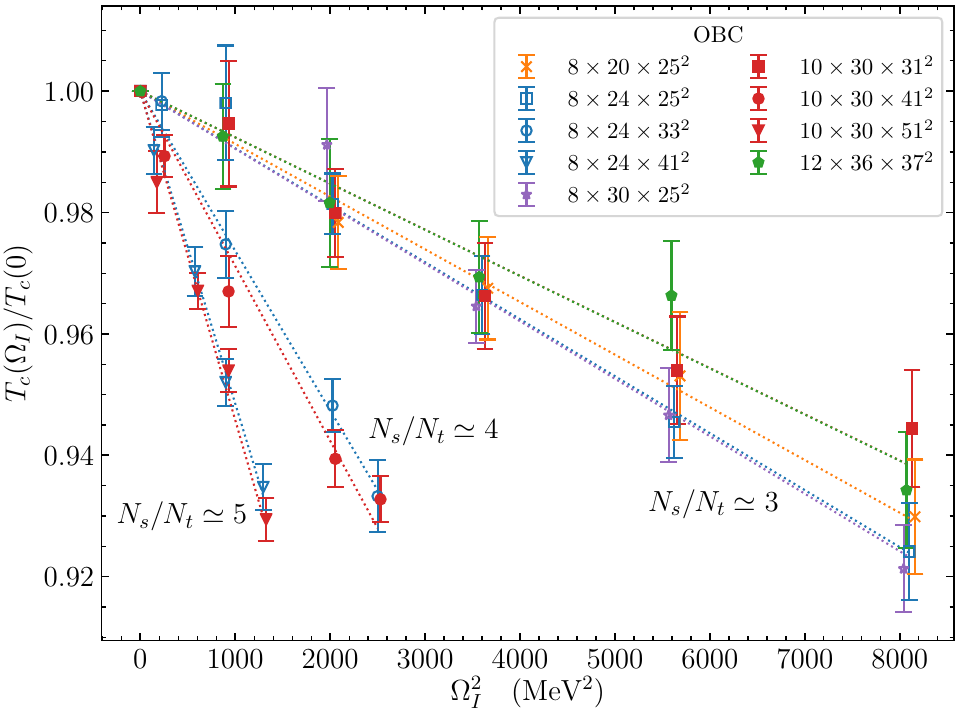}
}
\hfill
\subfigure[]{\label{fig:Tc-OBC-V}
\includegraphics[width=.48\textwidth]{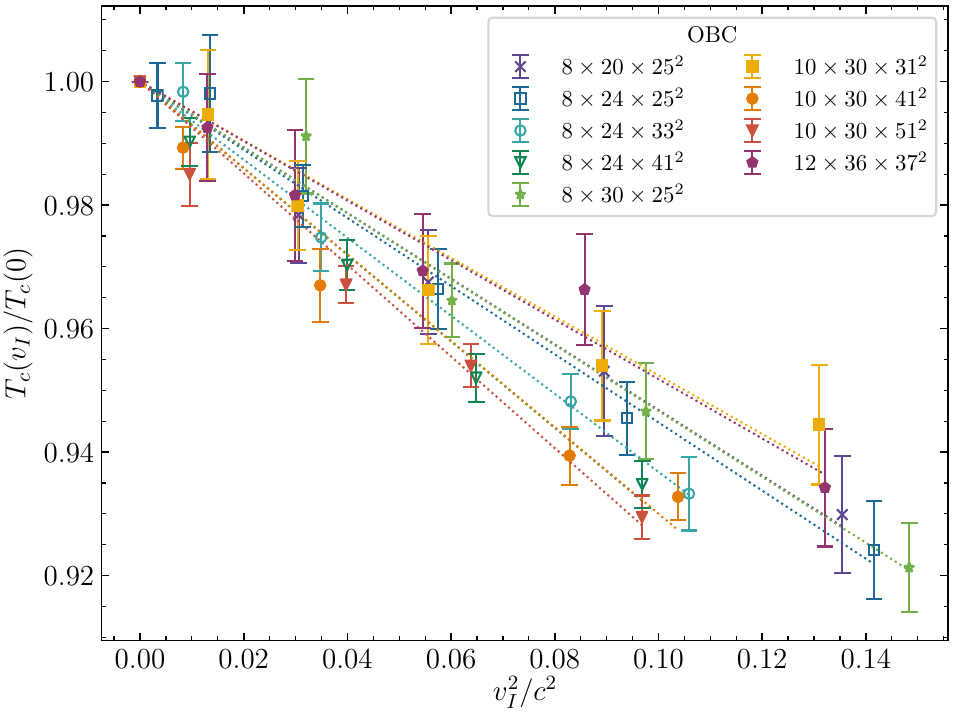}
}
\caption{The ratios $T_c/T_c(0)$ determined from the peak of the Polyakov loop susceptibility as a function of the imaginary angular velocity squared $\Omega_I^2$~\subref{fig:Tc-OBC-Om} and the linear boundary velocity squared $v_I^2$~\subref{fig:Tc-OBC-V}. Results are presented for several lattice sizes $N_t
\times N_z \times N_s^2$ with OBC. Lines correspond to simple quadratic fits $T_c(\Omega_I)/T_c(0)=1-C_2\Omega_I^2$ and $T_c(v_I)/T_c(0)=1-B_2 \,v_I^2/c^2$.  Fig. from~\cite{Braguta:2021jgn}.} 
\label{fig:Tc-OBC-var-all}
\end{figure}

The resulting dependence of the critical temperature on the $\Omega_I^2$ is presented in Fig.~\ref{fig:Tc-OBC-Om}. It can be clearly seen, that all points with the same aspect ratio $N_s/N_t$ lie almost on one straight line $T_c(\Omega_I)/T_c(0)=1-C_2 \Omega_I^2$. Quite remarkably, if we plot instead critical temperature versus the linear velocity on the boundary $v_I=\Omega_I (N_s-1)a/2$, then all points fall on one straight line $T_c(\Omega_I)/T_c(0)=1-B_2\, {v_I^2}/{c^2}$ (see Fig.~\ref{fig:Tc-OBC-V}). 

\begin{figure}[htb]
\begin{center}
\includegraphics[width=.6\textwidth]{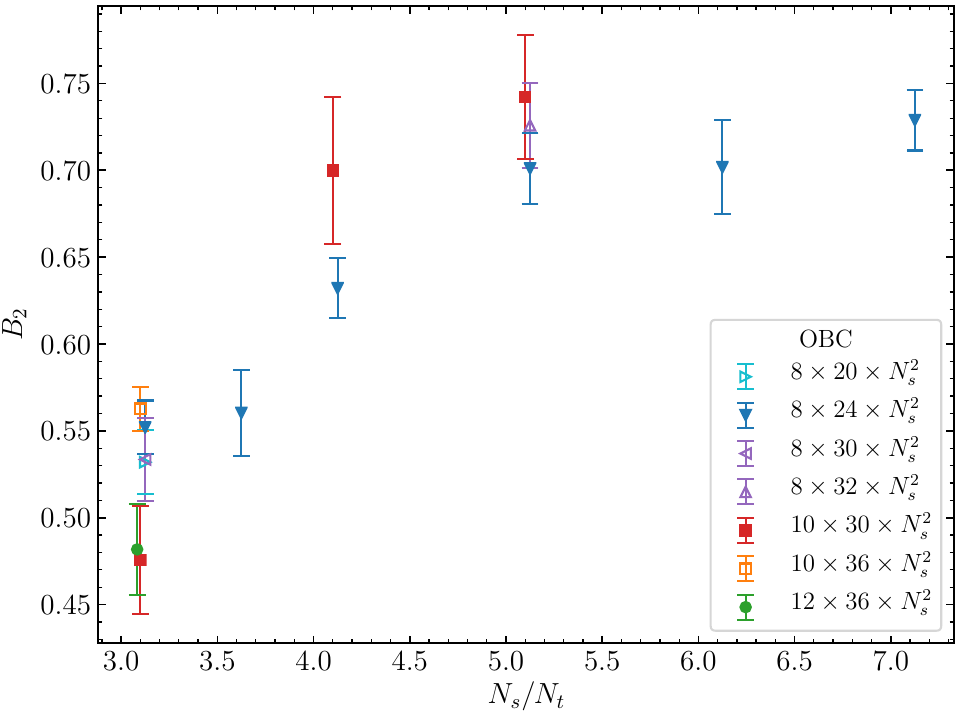}
\caption{The coefficient $B_2$ in Eq.~\eqref{eq:tempdep} as a function of the ratio $N_s/N_t$ for several lattice sizes with OBC. Fig. from~\cite{Braguta:2021jgn}.
}\label{fig:B2-OBC}
\end{center}
\end{figure}

Note that upon analytical continuation to real angular velocities these results transform into the following dependence:
\begin{equation}
\begin{split}
    \frac{T_c(\Omega)}{T_c(0)}=1+C_2 \Omega^2\, , \qquad C_2 > 0\, ,\\
    \frac{T_c(v)}{T_c(0)}=1+B_2 \frac{v^2}{c^2}\, , \qquad B_2 > 0\, ,
    \label{eq:tempdep}
\end{split}
\end{equation}
thus for real angular velocity the critical temperature of the confinement-deconfinement phase transition grows with the value of the angular velocity. In Fig.~\ref{fig:B2-OBC} the dependence of the coefficient $B_2$ on the aspect ratio is presented. It is almost independent on the lattice spacing and slightly increases with increasing lattice size in the transverse direction $N_s$. At sufficiently large values of $N_s$ the coefficient $B_2$ goes to a plateau $\sim 0.7$.

For other types of boundary conditions the behaviour of the system is very similar to OBC, the dependence can be described by the same simple formulae~\eqref{eq:tempdep}. 
At sufficiently large lattice sizes $N_s$ in transverse direction $B_2\sim 1.3$ for periodic boundary conditions and $B_2\sim 0.5$ for Dirichlet boundary conditions~\cite{Braguta:2021jgn}. Thus we conclude that growth of the critical temperature with the value of angular velocity is a universal property of rotating gluodynamics.

Note that Ehrenfest-Tolman effect would lead to decrease of the critical temperature with the angular velocity. It can be easily understood as follows. The space-dependent temperature, predicted by Ehrenfest-Tolman effect grows, when one moves from the rotational axis to the boundary. As the result, one needs lower temperature at the rotation axis in order for the system to be in the deconfinement. However we would like to stress, that our results are opposite: in gluodynamics the critical temperature grows with real angular velocity.

\section{Simulations with fermions}
Finally we present the first preliminary results of the phase diagram of the theory with fermions.
We used $N_f=2$ Wilson fermions for the simulation, (imaginary) rotation is introduced in the same way, as was done in~\cite{Yamamoto:2013zwa}. The simulation was performed on the lattice $4\times 20\times 21^2$ with OBC in $x$- and $y$- directions. The value of the hopping parameter is fixed at $\kappa=0.170$. The pion mass at $\beta=5.15$ is equal to $m_{\pi}=690$~MeV~\cite{Philipsen:2016hkv}.

\begin{figure}[htb]
\begin{center}
\includegraphics[width=.6\textwidth]{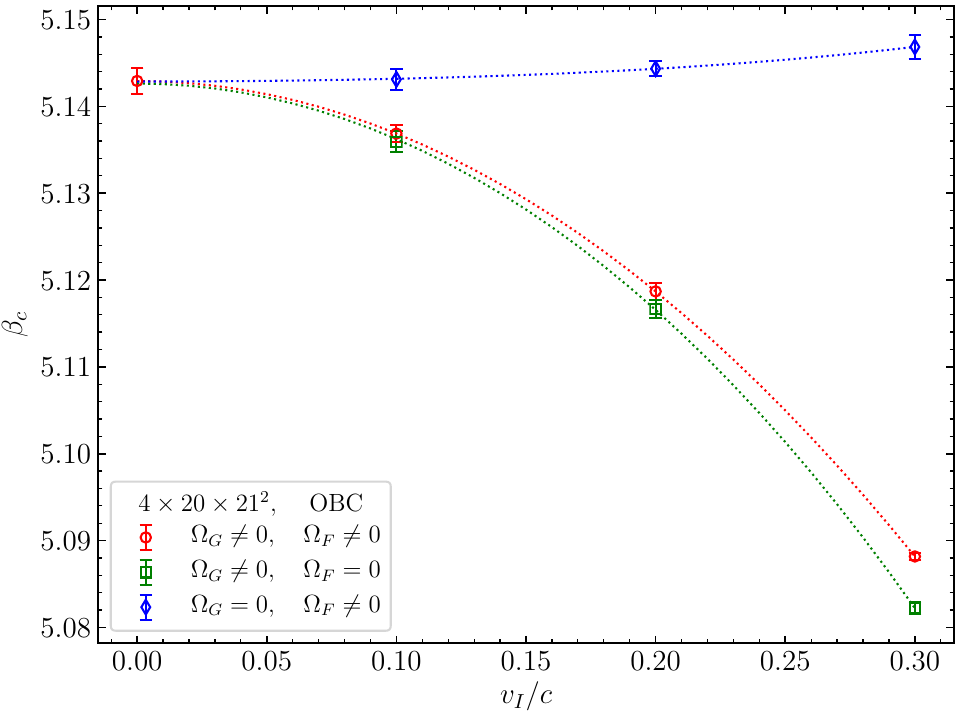}
\caption{The dependence of the confinement-deconfinement critical gauge coupling $\beta_c$ in the theory with $N_f=2$ fermions on the value of linear velocity at the boundary $v_I$ in the cases when separately fermionic and gluonic rotation is turned on and when full rotation is included. The result was obtained on the lattice $4\times 20\times 21^2$ with OBC, the hopping parameter is $\kappa = 0.170$. }\label{fig:TC-fermion}
\end{center}
\end{figure}

In this case apart from gluonic term, rotation also enters the fermionic part of the action: $S=S_G(\Omega) + S_F(\Omega)$.
In order to disentangle the effect of the rotation on fermions and gluons we introduced separate angular velocities for the gluonic and fermionic parts: $S=S_G(\Omega_G) + S_F(\Omega_F)$. Simulations have been performed in three cases: 1) when only fermionic contribution from the rotation is included ($\Omega_G=0, \Omega_F\neq 0$), 2) only gluonic contribution is included ($\Omega_G\neq 0, \Omega_F = 0$) and 3) when both effects are turned on ($\Omega_G=\Omega_F=\Omega_I\neq 0$).
The resulting dependence of the confinement-deconfinement critical gauge coupling $\beta_c$ on the velocity $v_I$ at the boundary is presented in Fig.~\ref{fig:TC-fermion}. It can be easily seen that if one includes only gluonic part of the rotation, then the critical $\beta_c$ and correspondingly the critical temperature decrease with increasing (imaginary) angular velocity, as was observed in gluodynamics. The effect of the rotation on fermions is completely opposite to a gluonic one. If one includes only rotating fermions, then the critical $\beta_c$ and critical temperature increase with the value of the imaginary angular velocity. The effect due to gluons is much higher and in the case, when the full interaction is turned on, the critical $\beta_c$ still decreases with $v_I$ as in gluodynamics, although fermions make this dependence somewhat milder.

After analytical continuation into real angular velocities $\Omega_I\to\Omega$ one gets, that rotating fermions decrease critical temperature, while rotating gluons increase critical temperature.
In total, rotation of both parts of the action together leads to increasing critical temperature. 
We also simulate the system with periodic boundary conditions and observe qualitatively the same behaviour. Notice, however, that one can expect the dependence of these results on the pion mass. So, it would be interesting to see, how this effect changes in the case of lighter pion masses. The same opposite effect of rotation on gluons and fermions was observed in \cite{Jiang:2021izj}, where the dependence of effective coupling on angular velocity is also taken into account within NJL model.

\section{Conclusions}
We have studied the phase diagram of gluodynamics with nonzero rotation by means of lattice simulations. In order to perform this study we switched into the rotating reference frame, in which rotation is reduced to external gravitational field. We considered several types of boundary conditions: open, periodic and Dirichlet. It was found that for any type of boundary conditions the dependence of the critical temperature on the value of the imaginary angular velocity is given by an expression: $T_c(\Omega_I)/T_c(0)=1-C_2\Omega_I^2$, where $C_2>0$. Coefficient $C_2$ depends on the size of the lattice in the transverse direction $N_s$, and we have found that this dependence can be made much milder, if one considers the linear velocity $v_I=\Omega_I {(N_s-1)} a/{2}$ at the boundary instead of angular velocity $\Omega_I$: $T_c(v_I)/T_c(0)=1-B_2\, v_I^2/c^2$. For a large lattices the coefficient $B_2$ is about $\sim0.7/1.3/0.5$ for open, periodic and Dirichlet boundary conditions. After the analytic continuation to real angular velocity this dependence transforms to: $T_c(\Omega)/T_c(0)=1+C_2\Omega^2$, thus one can conclude that {\it in gluodynamics critical temperature of the confinement-deconfinement phase transition grows with increasing (real) angular velocity}.

We have also presented the first lattice results for the critical temperature behaviour due to the rotation in theory with dynamical fermions.
We conclude that the effect of rotation on fermions and gluons is opposite: rotating gluons attempt to increase critical temperature, and at the same time rotating fermions try to decrease it. We plan to extend this investigation and to make more quantitative predictions with realistic pion masses.

Our results in gluodynamics are opposite to the arguments based on Ehrenfest-Tolman effect, which predict decreasing of the critical temperature due to the rotation from general relativity. It should be noted that various effective and phenomenological models also predict decreasing of critical temperature~\cite{Ebihara:2016fwa, Chernodub:2016kxh, Jiang:2016wvv, Zhang:2018ome, Wang:2018sur, Chernodub:2020qah, Chen:2020ath, Fujimoto:2021xix, Golubtsova:2021agl, Sadooghi:2021upd}.
It might indicate that the contribution from rotating gluons may be crucial for understanding the properties of rotating QCD medium.

\acknowledgments
This work was supported by RFBR grants 18-02-40126.
This work has been carried out using computing resources of the Federal collective usage center Complex for Simulation and Data Processing for Mega-science Facilities at NRC ``Kurchatov Institute'', \url{http://ckp.nrcki.ru/}  and the Supercomputer  ``Govorun'' of Joint Institute for Nuclear Research.

\bibliographystyle{JHEP}
\bibliography{biblio/biblio.bib}

\end{document}